\documentclass[a4paper, 12pt]{article}  
\title{Decuplets, glueballs and nonet anomalies}
\author{Micha\l\quad Majewski\\
Department of Physics and Applied Informatics \\
 University of Lodz\\
Pomorska 149/153, 90-236 Lodz, Poland\\
m.majewski@merlin.phys.uni.lodz.pl}

\usepackage{amsmath}
\begin{document}
\maketitle
\begin{abstract}


A new approach to problem of glueball search is presented. It refers to early J. 
Rosner's attempts to detect the tensor glueball. In the present description the 
glueball state is treated on equal footing with singlet $q\bar{q}$ one. Mixing 
glueball with $q\bar{q}$ nonet creates decuplet. Glueball can be detected as its 
component. Our approach is based on hypothesis of vanishing exotic commutators 
described as  VEC model. The model describes all multiplets of light mesons. This 
makes possible to compare the mass patterns of different multiplets. According 
to VEC description some abnormal nonets can be interpreted as incomplete decuplets. 
This makes possible to relate the anomaly of the nonet to a glueball component of 
decuplet. The model reflects rich diversity of strong interaction properties. The 
treatment presented is quite elementary: only masses of physical states are 
required to be known.   

\end{abstract}

\newpage
\section{The pure glueball meson is not necessary} 
In current opinion the quark-gluon picture \cite{Fri} well describes strong 
interactions. According to this picture the mesons are built out of quark (q) and 
antiquark ($\bar{q}$) which are coupled by gluon ($g$) exchange. It is supposed 
that this picture is valid for mesons having any signature $J^{PC}$ and in any 
mass region. This feature is described as \it{universality} 
\rm of the quark-gluon picture. The hypothetic $g$ is quark-less, flavor-less 
electrically neutral particle. It has the property of self-interaction which 
implies the existence of bound states of two or more gluons. Such an object is 
called glueball (G) and is a singlet state of SU(3) symmetry. Hence, G can 
interfere with  $q\bar{q}$ singlet and isoscalar octet states. 

The quark-gluon picture of strong interaction would be confirmed by the existence 
of G. This stimulates its experimental search. However, the hypothetic 
properties of G are too scanty for the needs of experimental investigation. The 
investigation turned out to be very difficult and during almost half a century 
did not provide satisfactory result. Desirable effect would be to 
detect a separate particle being pure G state. Such an object has not been 
found so far but during investigation considerable collection of particles 
which possibly are not $q\bar{q}$ states was discovered. These particles are not 
pure G but probably include its considerable component (see \cite{Clo} for 
the most recent reviews). The quest 
for G is still continued.

Although the pure G meson is not observed one cannot claim that G states do 
not exist. This may simply mean that the particles which are pure G states are 
unobservable. To be observable G must interact with other hadrons. For this to be 
the case 
it should have the ability to mix with larger unitary multiplet. Being mixed with 
nonet of $q\bar{q}$ states it forms a decuplet. Then it is subjected  
to restrictions imposed on decuplet components by SU(3) symmetry.  
G can be detected as one of the three interfering unphysical components creating 
the physical isoscalar states of decuplet. If it dominates one of the 
states then this state may be considered as the "G candidate". Hence, the 
existence of the G state can be established on the ground of unitary 
symmetry and observation of a pure G is not necessary. 

The unitary symmetry is perceptible due to the property of mesons to form 
multiplets which are collections of particles having different but definite flavors and 
the same signature $J^{PC}$. They form octets (O), nonets (N) and decuplets (D). 
Each  multiplet is described by some representation of unitary symmetry 
$SU(3)$. The $SU(3)$ multiplets are analogical to the $SU(2)$ ones describing 
electromagnetic interaction. However, the mass differentiation within SU(2) 
submultiplets of SU(3) multiplets is usually neglected, therefore, this 
symmetry is considered exact.  In the strong interactions  
the multiplets gather the particles having \it{a priori different} 
\rm masses. We call them "multiplets of broken $SU(3)$ symmetry". 
It is believed that the symmetry breaking announces the interaction. 

Several interactions can break $SU(3)$ symmetry. The most apparent effect 
of breaking is the difference between isotriplet and isodublet masses of the 
particles belonging to the same multiplet (e.g. $\pi$ and $K$ mesons). This 
difference is attributed to nonperturbative $g$ interaction and  cannot be 
calculated. However, the effect of this interaction can be described in 
the phenomenological approach by experimentally verifiable Gell-Mann - Okubo 
(GMO) formula for octet mesons. This formula fits the data and has the property 
of universality. We call this procedure the GMO-breaking
\footnote{$K-a$ determines all differences between masses of the octet 
states as $3(x_8-K)=K-a$}.

Other breakings are much weaker  and are masked by GMO one. They can be 
exhibited if the description of the multiplet complies with GMO requirement. 
The effect of such breaking is described as \it{anomaly} \rm of the multiplet 
shape. The anomalies can be observed in the flavor multiplets larger than octet. 
Anomalies of flavor symmetry comprise information on the unknown interactions 
which we want to investigate. They can be  observed through anomaly of multiplet 
mass pattern. They can be also seen due to pattern difference between two (or 
more) identical multiplets differing only by $J^{PC}$ signatures or belonging to 
various mass regions. Perhaps search for anomalies is the most promising way for 
detecting interactions.   

We begin with the question of how the anomalies can be recognized. In the next 
two sections we remind the VEC
description of the light meson (LM) multiplets  and 
define the benchmark multiplet which provides the pattern for anomalies search.

\section{VEC description of light meson  multiplets}
The approach refers to phenomenological investigations performed during the eighties 
of the twentieth century \cite{Ros}. They were attempting to detect G as an object 
causing deformation of $2^{++}$  and $0^{-+}$ nonet structures.
These attempts did not clarify much as they were premature. However, the very idea 
of such line of investigation cannot be questioned. The problem is in ability of 
its actual realization. Now we have much larger sample of data and more tools for 
their analysis. Therefore, it is probably a right time to come back to these ideas. 
\newpage
An opportunity for such an approach to be successful is provided by the VEC model  
\footnote{Model of vanishing exotic commutators (formerly described as ECM model 
\cite {TM})}  
which describes all multiplets of LM using the system of \it{master equations} \rm(ME) 
\cite{TM,MT,Loc}
\begin{equation} \label{A}
	\sum_{i=1}^n l_i^2x_i^r=\frac{1}{3}a^r +\frac{2}{3}b^r,  \quad  b\doteq 2K-a,  
\quad r=0,1,2,\ldots    
\end{equation}
where $r$ is power index. Particle symbols of $x_i$, $a$, $K$ 
stand for the mass squared of physical mesons, $l_i$ is the amplitude of octet 
content of isoscalar state $x_i$: 
\begin{equation}
|x_8\rangle=\sum_{i=11}^nl_i|x_i\rangle.    \label{C}
\end{equation}
Since isoscalar octet state $x_8$ and physical $x_i$ states describe uncharged 
states, the $l_i$ are real numbers, hence
\begin{equation}
l_i^2> 0, \qquad     i=1,2,\ldots n. \label{D}
\end{equation}   
The numbering of the isoscalar mesons is chosen so that $i$ 
increases with growing mass:   
\begin{equation}
x_i<x_{i+1}.  \label{E}
\end{equation}

The number of equations making ME system depends on multiplet. Solution of ME 
- the set of $l_i^2$'s - depends on parameters $a$, $b$, $x_i$. The structure 
of ME suggests that $x_i$ and ($a, b$) play different roles in description of 
multiplets: $x_i$ describes masses of individual isoscalar mesons while $a$ and $b$ 
have more complicated meaning and should be considered as known "theory 
constants". This is just what we need. If $a$ and $b$ are determined by 
measurement then GMO breaking requirement is approximately fulfilled and 
deformation of the multiplet exhibits anomaly.  

Also the structure of ME suggests that at fixed values of ($a$, $b$) the multiplets 
of several multiplicities are allowed. 

ME are linear with respect to unknown $l_i^2$'s. The solutions of such systems 
of equations are well known. We are looking for solutions which satisfy the 
positivity conditions (\ref{D}). The $l_i^2$'s are functions of masses 
$a, b, x_1,x_2,...$. Conditions (\ref{D}) which restrict these masses  help to 
test the affiliation of a given set of particles to the supposed multiplet. So 
they provide the criterion of relevant multiplet existence. Having known the 
positive solution $l_i^2$ of ME we can diagonalize the mass operator 
of the multiplet and determine its wave function. We thus describe broken SU(3) 
multiplet which is expressed in terms of physical masses \cite{Loc}. 

 

VEC predicts the existence of D \cite{Loc,Where}. The mesons 
$a$, $b$, $x_1$, $x_2$, $x_3$ belong to $D$ if they fit criteria  (\ref{D}). 
Then the wave function 
of $D$ can be constructed which components are determined by physical masses  
 of $D$ particles. The $G$ state can be distinguished and its 
unphysical mass can be fixed \cite{Loc}.

The determination of wave function requires very accurate data on masses. 
This is a merit of description, not its fault, as it determines the accuracy 
of predictions. However, excessive sensitivity can weaken predictive power 
of the procedure. Therefore, it is desirable to have also simpler criteria. 
They can be formulated within 
the VEC model as well. We explain below how this can be done but begin from 
presenting  some further features of ME which justify the procedure we 
propose.

\section{Varieties of flavor multiplets}
VEC predicts several multiplets which arise from solution of ME. The 
description of multiplets which include $n$ isoscalar mesons $x_i$ requires 
solving the ME with respect to unknown quantities $l_i^2$. To calculate the 
$l_i^2$'s we need the set of ME for $r=0, 1, \dots,(n-1)$. However, 
this is merely the minimal system of ME describing this multiplet.   

The same multiplet can be described by larger ME system provided this 
system satisfies some solvability conditions. We use a particular form of 
such conditions which is suggested by open structure of the ME set. We take 
into account subsequent ME for $r=n, n+1,\ldots$. The calculated $l_i^2$,   
expressed as the functions of the multiplet masses, should be inserted 
into these equations. It may happen that one of the equations (say, for $r=n$) 
is satisfied by these masses. Then this equation becomes the mass formula (MF) 
of the multiplet. Obviously, a multiplet may have more than one MF. 

The MF arises due to the restriction on the masses of $x_i$. Therefore, the 
number of MF cannot exceed $n$. The actual number $k$ of MF ($0\leq k\leq n$) 
should be determined from data fit for each multiplet separately. The 
number of ME to be considered for such a  multiplet is $n+k-1$.

The multiplets built on the same base $(a,b)$ and having the same $n$ but 
different $k$ are independent and have different patterns.  They are 
considered as \it{varieties} \rm of the same multiplet and marked 
by its currently used name indexed by $k$. The very existence of 
different varieties of the multiplet testifies the existence of various 
interactions influencing the structure of multiplet. 

There may exist three $N$ multiplets ($N_0, N_1, N_2$) 
and two $D$ multiplets $(D_0, D_1)$ which can be built out on the same 
$(a,b)$ basis. 
  

\section{Ideal nonet as a pattern}
Nonet arises due to the mixing of octet isoscalar state with an SU(3) 
singlet and is described as ME multiplet for $n=2$. Three old standing 
varieties of N are known: \\
 $N_0$ - known as Gell-Mann  - Okubo (GMO-nonet) having no MF\\
 $N_1$ - described as Schwinger (S-nonet) having one MF\\
 $N_2$ - ideal (I-nonet) having two MF \\ 
The I nonet is described by system of ME:
\begin{subequations}\label{F}
\begin{align} 
l^2_1+l^2_2&=1,\\   
l^2_1x_1+l^2_2x_2&=\frac{1}{3}a +\frac{2}{3}b,\\  
l^2_1x^2_1+l^2_2x^2_2&=\frac{1}{3}a^2 +\frac{2}{3}b^2,\\
l^2_1x^3_1+l^2_2x^3_2&=\frac{1}{3}a^3 +\frac{2}{3}b^3.
\end{align}
\end{subequations}
Solving the first two equations we calculate $l^2_1$, $l^2_2$ as the functions 
of masses. Next, substituting $l_i^2$'s into third and fourth equations 
we obtain two MF's which determine the masses of  mesons $x_1$, $x_2$. 
These MF's can be transfered to more familiar form:\\ 
\begin{equation} \label{G}
x_1=a, \quad  x_2=b,  \quad l^2_1=\frac{1}{3}, \quad l^2_2=\frac{2}{3}; \\ 
\quad |x_1> = |a>,  \quad |x_2> =|b>.
\end{equation}
where  $|a>$, $|b>$ are the nonet basic states:
\begin{equation} \label{H}
a=\frac{1}{\sqrt{2}}(u\bar{u}+d\bar{d}), \quad\ b=s\bar{s}. \\  
\end{equation}

This solution is determined by GMO breaking mechanism acting singly   
on the nonet states. It describes the shape of the nonet in a very simple 
way and is universal. Therefore, it can be used as a pattern for searching 
anomalies of flavor multiplets. 


\section{$D_1$ contents of the S-nonet components}
$\bullet$   $D$ includes three physical isoscalar states $x_i$ which can be 
represented as a superpositions of the base states $a$, $b$, $G$ 

The multiplet $D_1$ is determined by the solution $l_i^2$ (i=1,2,3) of the system of 
four ME 
\begin{subequations}
\begin{align}\label{J} 
l^2_1+l^2_2+l_3^2&=1,\\   
l^2_1x_1+l^2_2x_2+l^2_3x_3&=\frac{1}{3}a +\frac{2}{3}b,\\  
l^2_1x^2_1+l^2_2x^2_2+l_3^2x^2_3&=\frac{1}{3}a^2 +\frac{2}{3}b^2,\\
l^2_1x^3_1+l^2_2x^3_2+l_3^2x^3&=\frac{1}{3}a^3 +\frac{2}{3}b^3.
\end{align}
\end{subequations}
One can show that $D_1$ can be represented as superposition of $N_2$ and some SU(3) 
singlet. The nature of the singlet is undetermined. The VEC requirements accept all 
"extending singlets" which are usually mentioned like G, hybrid or multiquark state. 
The singlet G is favored as only this state is supposed to have the property 
of universality. The MF is: 
\begin{equation} \label{K}
(x_1-a)(x_2-a)(x_3-a)+2(x_1-b)(x_2-b)(x_3-b)=0.
\end{equation}
Combining (\ref{K}) with the criterion (\ref{D}) we find that the masses of $D_1$  
are subjected to further restrictions: the 
masses of the $D_1$ have to satisfy the mass ordering rule (MOR) \cite{Loc}
\begin{equation} \label{Ja}
x_1<a<x_2<b<x_3. \quad (MOR-D_1)
\end{equation}

MOR-$D_1$ divides accessible region of $x_i$ mesons into three isolated subregions 
which are separated by $a$ and $b$. 
In each of the subregions the states $x_i$ are uniformly dominated by $a$,  $G$, $b$.   
\begin{equation}  \label{Jb}
x_1 \sim   a,  \quad  x_2 \sim G, \quad x_3 \sim   b. 
\end{equation}
Therefore, it is convenient to introduce another notation
\begin{equation} \label{Jc}
x_a\doteq x_1.\quad x_G\doteq  x_2.\quad x_b\doteq  x_3.
\end{equation}  
which makes MOR-$D_1$ still more transparent: 
\begin{equation} \label{L}
x_a <a<x_G<b<x_b.
\end{equation}
$\bullet$  S-nonet ($N_1$) is considered to be a firmly established multiplet. 
It is announced for many $J^{PC}$ mesons and dominates the perception of LM 
spectroscopy. However, the constituents of its isoscalar components and diversity 
of S-nonet shapes remain vague. We argue that these problems arise from G mixing. 

$N_1$ is described by the first three equations of the system (\ref{F}). Its MF is 
\begin{equation} \label{I}
(a-x_1)(a-x_2)+2(b-x_1)(b-x_2)=0.
\end{equation}
As the S-nonet has one MF we need an extra information on the masses of $x_1$, $x_2$ 
mesons for evaluating  $l_1^2$, $l_2^2$. One can use for that purpose the known 
value of one of the masses and calculate the other one with the help of MF. We 
can see that the pair of masses determined this way is different from the values 
of masses of the I-nonet (\ref{G}). The change of S-nonet masses $x_1$, $x_2$ 
relative to the I-nonet ones shows the anomaly. It is thus compatible with the 
existence of an extra state.

The components of $N_1$ have to comply with one of the two MOR conditions \cite{In}
\begin{subequations}\label{H}
\begin{align}
\label{Ha}
(a) \quad a<x_1<b<x_2, \quad   (MOR-N_{1(a)}),\\
\label{Hb}
(b) \quad x_1<a<x_2<b.  \quad  (MOR-N_{1(b)}).  
\end{align}
\end{subequations}
These conditions determine two completely different nonets which we describe 
as $N_{1(a)}$  and $N_{1(b)}$ ones. Both of them are observed
\footnote{The current description of $N_1$ as S-nonet uses the  
mixing angle $\vartheta$ for determination the isoscalar states. The allowed 
regions of  $\vartheta$ are:  
for (\ref{Ha}) $\tan^2\vartheta>\tan^2\vartheta^{id}$ and for (\ref{Hb}) 
$\tan^2\vartheta<\tan^2\vartheta^{id}$,    
where $\vartheta^{id}=35.26^o$ is ideal mixing angle \cite{Sum}.}

If $N_1$ is built out on the same ($a$, $b$) base as $D_1$ then the MOR's 
(\ref{H}) may be considered as incomplete MOR-$D_1$ (\ref{L}). The comparison 
shows that\\
--- if $x_1,x_2\in N_{1(a)}$ then they are dominated by ($G,b$) components of $D_1$, \\
--- if $x_1,x_2\in N_{1(b)}$ then they are dominated by ($a,G$) components of $D_1$, \\
respectively. Both types of the $N_1$ include $G$ state. Therefore, the very 
existence of $N_1$ justifies the existence of $G$ which can be only seen  
as the state of $D_1$. This suggests that it plays an essential role in the  
structures of $D_1$. Perhaps within this multiplet the suitable $G$ is always 
"ready for use" since it is built of the  gluons which mediate interactions 
between quarks which are present there. This is the way $G$ becomes a  
constituent of the isoscalar mesons of $D_1$. 

The current description of N does not explain the origin of the S-nonet 
anomalies. Moreover, N themselves are distinguished by the results of biased 
experiments ignoring the possibility of D appearance. Perhaps the results 
of these measurements should be reanalyzed. Also extending these 
experiments and increasing their accuracy is necessary. 
It is possible that the nature of these multiplets has been for a long 
time misunderstood. The explanation of this confusion may have far-reaching 
implications for meson spectroscopy. Some of the implications can be seen 
immediately.
\newpage

\section{Unrecognised glueballs and missing mesons}
$\bullet$  
We have established that all S-nonets include $x_G$ dominated state.
Reviewing the particle data \cite{RPP} we find that the mesons  
\begin{equation} \label{N}
f_1(1285),\quad h_1(1380),\quad\ \eta(1405),\quad f_2(1430)
\end{equation}  
should be G dominated. The decay modes of $f_1(1285)$ and $h_1(1380)$ 
do not contradict these assignments; the G dominated structure of 
$\eta(1405)$ meson established earlier \cite{FF,Where} is now  confirmed; 
the $f_2(1430)$ should have G dominated structure if it exists \cite{Loc}. \\
$\bullet$  
The old standing puzzle of exceptional properties of $f_1(1260)$  
\begin{equation} \label{O}
m=1230\pm 40 MeV, \quad \Gamma =250\div 600 MeV   \\
\end{equation}
is solved by changing its affiliation from $N_1$ to $D_1$. 
If this signal belongs to $D_1$ then it should be attributed to two different 
particles: isosinglet meson $x_a$ and isotriplet meson $a_1$: 
\begin{equation}  \label{P}
x_a,  \quad   \quad  a_1
\end{equation}
which have similar modes of decay. \\
$\bullet$    The observed axial-vector mesons are 
collected into the $N_1$ multiplets having $J^{PC}$=$1^{++}$ and $J^{PC}$=$1^{+-}$ (described 
as $N_{1A}$ and $N_{1B}$), where instead of the physical $K_1$(1270) 
and  $K_1$(1400) there stand their C-even or C-odd combinations:
\begin{subequations}  \label{Q}
\begin{align}
K_{1A}=K_1(1270)cos\phi -K_1(1400)sin\phi, \\
K_{1B}=K_1(1270)sin\phi +K_1(1400)cos\phi  
\end{align}
\end{subequations}
Joint MF's analysis of data on $N_{1A}$ and $N_{1B}$ gives the following 
values for these masses \cite{Sum}  
\begin{subequations}  \label{Q}
\begin{align}
K_{1A}=(1340\pm 8)^2 MeV^2,\\
K_{1B}=(1324\pm 8)^2 MeV^2.
\end{align}
\end{subequations}
and appoints the nonets $N_{1A}$ and $N_{1B}$ as $N_{1(a)}$ and $N_{1(b)}$ 
respectively. Consequently, we define also $D_1$ for these mesons as $D_{1A}$ 
and $D_{1B}$. Observe that the slight difference between "bare" $K_{1A}$ 
and $K_{1B}$ masses is strongly amplified by hadronic interactions.  

The states of the $N_{1A}$ and $N_{1B}$ are not comparable, but the states of 
$D_{1A}$ and $D_{1B}$ can be compared. 
The fact that basic masses (a,b) of $D_{1A}$ and $D_{1B}$  are not identical 
can only increase interest to this comparison because it demonstrates the
dependence of $D_1$ properties  on  C-parity. Hence, they disclose 
the influence of weak interaction on meson multiplets.
(cf. $K_S$ and $K_L$ states of pseudoscalar mesons). 
\section{Call for new data}
Anomalies of the S-nonets provide the evidence for the existence of further (beyond GMO) 
mechanisms of SU(3) breaking. The anomalies are caused by interactions which are unknown. 
It is just a purpose of the S-nonet investigation to recognize their nature. 
The anomalies provide much weaker breaking than the GMO one \cite{Loc}. This requires 
much more accurate data to make them observable.
The present data (partly old and skimpy) enable us to select few nonets but  
are insufficient for completing decuplets. Therefore, for the sake of 
present and future development of meson spectroscopy it is necessary 
to increase accuracy of the data and extend the measurements to other $J^{PC}$ mesons.

\section{Acknowledgments}
Author thanks Professor Anna Urbaniak-Kucharczyk - Dean of the Physical 
Department of the University of Lodz; Professor  Pawe\l{} Ma\'slanka and 
Professor Jakub Rembieli\'nski - Chiefs of Theoretical Departments for 
their support; Special thanks are 
expressed to Professor Piotr Kosi\'nski for many interesting discussions,
valuable comments and reading the manuscript; Dr Bartosz Zieli\'nski's help 
in computer manipulations is grateful appreciated.

\section*{Afterword\\ \large The rank of the VEC model description}

$\bullet$ The paper is a further report on the progress in analyzing  flavor 
aspects of light mesons. Project of this research program arose from 
curiosity whether strong interactions can be investigated by traditional 
three-step methodology of natural sciences:\\
\begin{displaymath}
 \bf data   \longrightarrow  \bf model \longrightarrow  \bf theory \\
\end{displaymath}
where\\
"data" are measured masses and quantum numbers of the particles, \\
"model" is a model describing particle multiplets, \\
"theory" is QCD.\\ 
The phenomenological model describing data should be neutral (not intentional).
The approaches inspired by a theory (or doctrine) cannot be used as they 
are purposed to confirm some hypothesis formulated before. 
We adopt the VEC model which describes all multiplets 
and unifies their description. The main tool of the VEC investigation of LM is provided 
by set of master equations (ME) which describe the multiplets of broken SU(3) 
symmetry in terms of masses. The most significant results, among those already published, 
are: \\
-- mass criterion for recruitment particles to the multiplet, \\
-- derivation of the mass formulas (MF),\\
-- discovery of the mass ordering rules (MOR), \\ 
-- demonstration that ideal mixing follows from ME,\\
-- prediction of the decuplet ($D$) existence and constructing its wave function, \\
-- confirmation of the glueball ($G$)-dominated structure of $\eta$ (1405) meson, \\
-- determination of $G$ localization within $D$,\\
-- separation of the multiplet "varieties" (discussed in this paper).\\ 
These relations confirm and extend current knowledge about the properties of broken 
SU(3) symmetry multiplets of LM and exhibit the capabilities of VEC model. Moreover, 
they are universal (hold  for all $J^{PC}$).\\ 
However, the key advantage of the VEC model is its ability to determine the "multiplet 
anomaly" and to relate it with an extra interaction. Knowledge of the listed  relations 
is sufficient for confirming the existence of $G$ and indicating the $G$-dominated 
states of $D_1$ for different $J^{PC}$. 


\newpage

$\bullet$ The main purpose of the present paper is to recognize the "habitation" of the 
$G$ states in the range of LM multiplets. The problem is explained in the 
article. Here we are  limiting ourselves to some rather loose remarks. 

Unsuccessful attempts to discover the pure G support the idea that G-state 
is hidden in the internal structure of D; therefore, the latter structure should 
be analyzed in more detail.  A suitable basis for the analysis of multiplets 
including $G$ is provided by the states $a$ and $b$. In this basis both the 
$N_1$ and $D_1$ multiplets can be described. If they are based on the same 
($a$, $b$) then they are correlated. In particular, it is possible to 
introduce some form patterns  ("benchmarks") which help to detect anomalies. 
Another sign of correlation can be seen by comparison of the MOR's of $N_1$ and $D_1$ 
multiplets. Each of two possible MOR sequences for  $N_1$ can be viewed as 
incomplete MOR sequence for $D_1$ cut on one of the ends. As $G$ resides in 
the center of MOR sequence for $D_1$ each observed $N_1$ includes $G$ as 
a component of one of its isoscalar states. 
Moreover, as MOR for $D_1$ is an universal relation, it follows that all 
observed $N_1$ multiplets are habitats of $G$.  Universality of $D_1$ MOR 
justifies also hypothetic list (16) of $G$-dominated mesons as well as 
speculations concerning the situation in the $a_1$ region. 



A comment is due to "theory" -- the last term of the three-step methodology 
of natural investigation. The "theory" should explain the nature of the object under 
study and provide its phenomenological description. QCD, being the only recognized 
candidate to this role, fulfills former of these requirements but  
does not satisfy the latter. This is currently 
explained as an effect of nonperturbative  properties of QCD in the LM region. 
The VEC analysis supports this interpretation. 

So the theory QCD and the model VEC belong to different levels of description.
They use different notions and terminology. QCD predicts theoretically 
motivated interactions while VEC distinguishes those which are the most 
visible. 


\end{document}